\documentclass[aps,showpacs,showkeys,twocolumn,twoside]{revtex4}

\usepackage{graphicx,eepic,epic,epsfig,amsmath,latexsym,amssymb,psfrag}

\usepackage{verbatim}

\usepackage{theorem} 

\newtheorem{definition}{Definition}

\newtheorem{lemma}[definition]{Lemma}

\def\squareforqed{\hbox{\rlap{$\sqcap$}$\sqcup$}}
\def\qed{\ifmmode\squareforqed\else{\unskip\nobreak\hfil
\penalty50\hskip1em\null\nobreak\hfil\squareforqed
\parfillskip=0pt\finalhyphendemerits=0\endgraf}\fi}
\def\endenv{\ifmmode\;\else{\unskip\nobreak\hfil
\penalty50\hskip1em\null\nobreak\hfil\;
\parfillskip=0pt\finalhyphendemerits=0\endgraf}\fi}

\mathchardef\ordinarycolon\mathcode`\:
\mathcode`\:=\string"8000
\def\vcentcolon{\mathrel{\mathop\ordinarycolon}}
\begingroup \catcode`\:=\active
  \lowercase{\endgroup
  \let :\vcentcolon
  }

\newcommand{\captionfonts}{\small}
\makeatletter  
\long\def\@makecaption#1#2{%
  \vskip\abovecaptionskip
  \sbox\@tempboxa{{\captionfonts \textbf{#1}\; #2}}%
  \ifdim \wd\@tempboxa >\hsize
    {\captionfonts \textbf{#1}\; #2\par}
  \else
    \hbox to\hsize{\hfil\box\@tempboxa\hfil}%
  \fi
  \vskip\belowcaptionskip}
\makeatother   

\newcommand{\nc}{\newcommand}
\nc{\rnc}{\renewcommand}
\nc{\beq}{\begin{equation}}
\nc{\eeq}{{\end{equation}}}
\nc{\beqa}{\begin{eqnarray}}
\nc{\eeqa}{\end{eqnarray}}
\nc{\lbar}[1]{\overline{#1}}
\nc{\bra}[1]{\langle#1|}
\nc{\ket}[1]{|#1\rangle}
\nc{\ketbra}[2]{|#1\rangle\!\langle#2|}
\nc{\braket}[2]{\langle#1|#2\rangle}
\nc{\proj}[1]{| #1\rangle\!\langle #1 |}
\nc{\avg}[1]{\langle#1\rangle}
\rnc{\max}{\operatorname{max}}
\nc{\Rank}{\operatorname{Rank}}
\nc{\smfrac}[2]{\mbox{$\frac{#1}{#2}$}}
\nc{\Tr}{\operatorname{Tr}}
\nc{\ox}{\otimes}
\nc{\dg}{\dagger}
\nc{\dn}{\downarrow}
\nc{\cA}{{\cal A}}
\nc{\cB}{{\cal B}}
\nc{\cC}{{\cal C}}
\nc{\cD}{{\cal D}}
\nc{\cE}{{\cal E}}
\nc{\cF}{{\cal F}}
\nc{\cG}{{\cal G}}
\nc{\cH}{{\cal H}}
\nc{\cI}{{\cal I}}
\nc{\cJ}{{\cal J}}
\nc{\cK}{{\cal K}}
\nc{\cL}{{\cal L}}
\nc{\cM}{{\cal M}}
\nc{\cN}{{\cal N}}
\nc{\cO}{{\cal O}}
\nc{\cP}{{\cal P}}
\nc{\cR}{{\cal R}}
\nc{\cS}{{\cal S}}
\nc{\cT}{{\cal T}}
\nc{\cU}{{\cal U}}
\nc{\cX}{{\cal X}}
\nc{\cY}{{\cal Y}}
\nc{\cZ}{{\cal Z}}
\nc{\csupp}{{\operatorname{csupp}}}
\nc{\qsupp}{{\operatorname{qsupp}}}
\nc{\rar}{\rightarrow}
\nc{\lrar}{\longrightarrow}

\def\e{\epsilon}

\def\Ph{\Phi}

\nc{\RR}{{{\mathbb R}}}
\nc{\CC}{{{\mathbb C}}}
\nc{\FF}{{{\mathbb F}}}
\nc{\NN}{{{\mathbb N}}}
\nc{\ZZ}{{{\mathbb Z}}}
\nc{\PP}{{{\mathbb P}}}
\nc{\QQ}{{{\mathbb Q}}}
\nc{\UU}{{{\mathbb U}}}
\nc{\EE}{{{\mathbb E}}}

\nc{\id}{{\mathbb I}}

\def\ot{\otimes}
\newcommand{\eq}[1]{Eq.~(\ref{eq:#1})}

\nc{\be}{\begin{equation}}
\nc{\ee}{{\end{equation}}}
\nc{\bea}{\begin{eqnarray}}
\nc{\eea}{\end{eqnarray}}
\nc{\<}{\langle}
\rnc{\>}{\rangle}
\nc{\Hom}[2]{\mbox{Hom}(\CC^{#1},\CC^{#2})}
\nc{\rU}{\mbox{U}}

\def\lpm{ \left(\rule{0pt}{2.1ex}\right. \!}
\def\rpm{ \!\left.\rule{0pt}{2.1ex}\right) }
\def\lbL{ \left[\rule{0pt}{2.4ex}\right. \!}
\def\rbL{ \!\left.\rule{0pt}{2.4ex}\right] }
\def\lpL{ \left(\rule{0pt}{2.4ex}\right.\!\!}
\def\rpL{ \!\! \left.\rule{0pt}{2.4ex}\right)}
\def\lpH{ \left(\rule{0pt}{3.0ex}\right.\!\!}
\def\rpH{ \!\! \left.\rule{0pt}{3.0ex}\right)}

\nc{\ob}[1]{#1}
\def\non{\nonumber}
\rnc\ss{\hspace*{0.1ex}}
\nc\ms{\hspace*{-0.1ex}}

\nc{\mPr}[1]{
	\begin{array}{c}
	\rule{0pt}{3.0ex}\mbox{\rm Pr} \\ \raisebox{1ex}{\scriptsize $#1$}
	\end{array} \!\!
}

\begin{document}

\title{Superdense coding of quantum states}

\author{Aram Harrow}
\affiliation{MIT Physics Dept., 77 Massachusetts Avenue, Cambridge, 
MA 02139, USA}
\author{Patrick Hayden}
\affiliation{Institute for Quantum Information, Caltech 107--81, Pasadena, 
CA 91125, USA}
\author{Debbie Leung}
\affiliation{Institute for Quantum Information, Caltech 107--81, Pasadena, 
CA 91125, USA}

\date{\rule{0pt}{3.2ex}\today}


\begin{abstract}
We describe a method for {\em non-obliviously} communicating a $2l$-qubit
quantum state by physically transmitting $l + o(l)$ qubits, and by
consuming $l$ ebits of entanglement plus some shared random bits.
In the non-oblivious scenario, the sender has a classical description
of the state to be communicated.
Our method can be used to communicate states that are pure or
entangled with the sender's system; $l + o(l)$ and $3l + o(l)$ shared
random bits are sufficient respectively.

\end{abstract}



\maketitle


\raggedbottom
\parskip .5ex
\marginparsep -0.25in
\marginparwidth 0.75in


%
{\bf Introduction.~~}
One of the most striking effects in quantum information theory is
known as superdense coding~\cite{Bennett92}. By making use of shared
entanglement, it is possible to communicate classical information at
twice the rate one would na\"{i}vely expect is allowed by causality. That
is, by physically transmitting only one qubit (a two-level system such as
a spin-$1/2$ particle) while at the same time consuming one ebit (the
shared state $(\ket{00}+\ket{11})/\sqrt{2}$), it is possible to
communicate two classical bits worth of information.  This observation
can be summarized by the following schematic inequality:
\begin{equation*}
1 \; \mbox{qubit} + 1 \; \mbox{ebit} \succeq 2 \; \mbox{cbits}.
\end{equation*}

%
It is natural to ask whether it is possible, using the same resources,
to communicate two \emph{qubits} worth of quantum information rather
than just two classical bits.
A simple thought experiment reveals that this should not be the case.
Indeed, if the schematic inequality
\begin{equation*}
1 \; \mbox{qubit} + 1 \; \mbox{ebit} \succeq 2 \; \mbox{qubits}
\end{equation*}
were true, then the two qubits communicated using just a qubit and an
ebit could themselves be paired with two ebits, resulting in the
communication of four qubits worth of quantum information.  Repeating
the process, an arbitrary amount of quantum information could be
transmitted by sending just the single original qubit and a
correspondingly large amount of entanglement.  This is known to be 
impossible \cite{Holevo73d}. 
For similar reasons, entanglement cannot increase the quantum capacity
of a noiseless quantum channel.
This leads to a strong dichotomy: 
entanglement can double the classical communication
capacity of a noiseless quantum channel but does not increase the
quantum communication capacity at all.

The argument for the latter claim rests on an important assumption about
quantum communication, however -- that the sender (Alice) {\em forwards} or
{\em delivers} a quantum state without knowing what it is, a condition
known as ``oblivious encoding.''
Equivalently, Alice's action is required to be independent of the 
transmitted state. 
Oblivious quantum communication automatically preserves entanglement
between the transmitted state and any other system.
In fact, recursive use of superdense coding as described in the previous
paragraph requires that it preserve 
entanglement between the transmitted and the receiver's (Bob's)
systems.  

Thus, the situation can be very different if Alice is given a
classical description of the state to be communicated and can alter
her encoding operation accordingly -- we call such encoding
``non-oblivious.''  An important example of non-oblivious encoding is
when Alice chooses the communicated state herself, as is the case,
for instance, with quantum digital signatures~\cite{Gottesman01}.
Non-oblivious encoding can be more powerful than the oblivious
version.  This is true, for example, in remote state 
preparation~\cite{Lo00rsp}, the variant of 
teleportation~\cite{Bennett93} in which
the sender knows the state to be communicated.  Remote state
preparation requires only half of the communication resources used in
teleportation~\cite{Bennett03}.

In this paper, we focus on ``non-oblivious'' communication of quantum
states using the dual resources of quantum communication and
entanglement.  
This is analogous to remote state preparation but the
classical communication is now replaced by quantum communication.  We
restrict our discussion to the tasks of ``preparing'' a pure state in
Bob's system or ``sharing'' a pure state that is entangled between
Alice's and Bob's systems. (We avoid the term ``transmitted state,''
which can be confused with the system that is physically transmitted.)
Our goal is to find out the extent to which entanglement can improve the
quantum communication capacity of a noiseless quantum channel in the
non-oblivious scenario.

{\bf Statement of result.~~} Our main result is that an arbitrary
$2l$-qubit state can be prepared or shared with high probability if
Alice transmits $l + o(l)$ qubits to Bob and if they share $l$ ebits
and at most $O(l)$ random classical bits.  No shared randomness is
needed in the important cases of communicating tensor product states
or arbitrary states drawn according to a known probability
distribution.  Our result is a generalization of superdense coding to
quantum states in the asymptotic and non-oblivious scenario.  The
possibility of superdense coding hinges on non-obliviousness.  By
Holevo's theorem~\cite{Holevo73d}, both versions of superdense coding
use a minimal amount of communication, a rate at which the
entanglement cost is also optimal.

In the following, we first discuss the task of preparing pure states,
starting with a protocol that requires no shared randomness but only
succeeds with some potentially small probability.  Then, we describe a slight
modification that uses shared randomness to ensure high probability of
success for all states.  Finally, we describe a generalization to
share pure entangled states between Alice and Bob.


{\bf An exact probabilistic protocol.}~~ 
We first describe a protocol for Alice to prepare any
$d^2$-dimensional state $|\psi\>$ in Bob's system, by sending $\log d$
qubits and consuming $\log d$ ebits of shared entanglement.
The protocol succeeds with a probability that depends on the state
prepared.
Fix a basis $\{\ket{i}_A\ket{j}_B\}_{1\leq i,j \leq d}$ for
$\CC^d\otimes \CC^d$ (or equivalently $\CC^{d^2}$).  
Alice and Bob initially share $\log d$ ebits, or equivalently 
the maximally entangled state $\ket{\Ph_d} = \smfrac{1}{\sqrt{d}}
\sum_{i=1}^d \ket{i}_A \ket{i}_B$.
The state to be prepared, $|\psi\>$, can always be written as
\bea
	\ket{\psi} 
	= {1 \over \sqrt{d}} \sum_{i,j} x_{i,j} \ket{i}_A \ket{j}_B
	= (X \otimes I) \ket{\Ph_d} 
\label{eq:mestrick}
\eea
where $X :=\sum_{i,j} x_{i,j} \, \ket{i}\bra{j}$.  The identity 
\bea
	\rho_B := \Tr_A \ket{\psi}\bra{\psi} = {1 \over d} X^T X^*
\label{eq:rhob}
\eea
will be useful later.  
\eq{mestrick} provides a simple scheme to prepare the state $|\psi\>$
-- Alice applies $X$ to her half of $|\Phi_d\>$ and then sends it to
Bob:
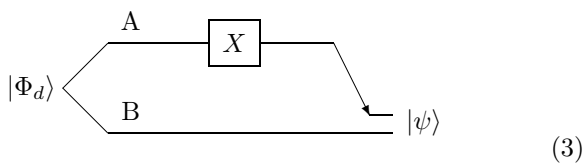
\bea
\setlength{\unitlength}{0.6mm}
\centering
\begin{picture}(90,35)
\put(-7,10){\makebox(10,10){$|\Phi_d\>$}}
\put(5,15){\line(1,1){10}}
\put(5,15){\line(1,-1){10}}
\put(15,25){\makebox(10,10){A}}
\put(15,5){\makebox(10,10){B}}
\put(15,25){\line(1,0){10}}
\put(15,5){\line(1,0){63}}
\put(25,25){\line(1,0){10}}
\put(35,25){\line(1,0){2.5}}
\put(37.5,20){\framebox(11,10){$X$}}
\put(48.5,25){\line(1,0){2.5}}
\put(51,25){\line(1,0){10}}
\put(60,25){\line(1,0){5}}
\put(65,25){\vector(1,-2){8}}
\put(73,9){\line(1,0){5}}
\put(80,2.5){\makebox(10,10){$|\psi\>$}}
\end{picture}
\label{eq:prprotocol}
\eea
In the above circuit and throughout the paper, time goes from left to
right, single lines represent quantum registers, and registers
connected in the left are initially in a maximally entangled state.
The arrow in \eq{prprotocol} represents a register sent from Alice to
Bob.

This scheme only succeeds if $X$ is applied successfully, which 
only occurs with some probability, because $X$ may not
be unitary.
One way to perform $X$ is via a generalized measurement~\cite{Peres95}
$\rho\rightarrow \sum_k E_k\rho E_k^\dag$ with Kraus operators 
\bea
	E_0 = \frac{X}{\|X\|_\infty} \,,~~~
	E_1 = \sqrt{I-E_0^\dag E_0}
\eea
where the operator norm of $X$, $\|X\|_\infty$, can be taken to be square
root of the largest eigenvalue of $X^\dag X$.  
Then, $E_0^\dag E_0 \leq I$ and the measurement is well-defined.
When the measurement outcome is $0$, $X$ is successfully applied, 
and this occurs with probability 
\bea
	\Tr E_0^\dag E_0 \, \smfrac{I}{d} =  
	{\! \Tr X^\dag X \over \; d \; \|X^\dag X\|_\infty} 
	= {1 \over \; d \; \|\rho_B\|_\infty} \geq {1 \over {1+\e}} \,. 
\label{eq:prbdd}
\eea
We have used \eq{rhob} to obtain the equalities in \eq{prbdd}.  The
parameter, $\e$, is defined by $\| \ss \rho_B \|_\infty \leq
\smfrac{1}{d} (1+\e)$.  It measures the deviation of the state 
$|\psi\>$ from being maximally entangled.  
This probability of success ranges from $1$ for a maximally entangled
state ($\e=0$) to $1/d$ for a product state ($\e = d-1$).
This implementation of $X$ requires the measurement outcome be sent to
Bob, but it suffices to use one extra qubit of communication, which is
negligible for large $d$.

\vspace*{1ex}


{\bf High probability protocol for arbitrary pure states.}~~ If
$\ket{\psi}$ is chosen randomly, then with high probability it will be
highly entangled ($\e$ is small)
\cite{Lubkin78,Lloyd88,Page93} and thus we would expect the protocol 
described above to succeed with high probability.

Suppose our goal is to find a protocol that succeeds with high
probability for any choice of input state $\ket{\psi}$, including
product states.  The above protocol can be adapted easily if Alice and
Bob share correlated random bits --
Alice will instead prepare $U\ket{\psi}$, with $U$ chosen according to
random bits shared with Bob.  Then, Bob can undo $U$ after receiving
$U|\psi\>$.  With high probability, the totally random $U|\psi\>$ is
highly entangled so that the probabilistic protocol would succeed with
high probability.

To make this intuition precise, we will analyze how much randomness is
required to ensure a given probability of success.  
The answer is provided by a lemma that is proved in
the Appendix:
\begin{lemma} \label{lem:pure}
Let $0 < \e \leq 1$.  If $d \geq \smfrac{10}{\e}$, there exists a set
of isometries $\{U_k\}_{k=1}^{n}$, where $n = \smfrac{120 \ln
2}{\e^3} \, d \log d$, such that
\bea
         \forall |\psi\>, ~\mPr{k} 
	 \! \left( 
 	 \| \Tr_A U_k |\psi\>\<\psi| U_k^\dagger \, \|_\infty \! < \!
         \smfrac{1+\epsilon}{d} \right) \geq  1\!-\!\e \,.
\label{eq:long}
\eea
Here, each $U_k$ takes $d^2$-dimensional states into a Hilbert space
${\cal H}_A \ot {\cal H}_B$, where $\dim(\cH_A) =
\smfrac{112 \ln 2}{\e^2} \, d \log d$ and $\dim(\cH_B) = d$.
\end{lemma}
Our lemma states that for any state $|\psi\>$, choosing $U_k$ randomly
out of $n = \smfrac{120 \ln 2}{\e^3} \, d \log d$ possibilities will
guarantee, with
probability at least $1-\e$, that the state $U_k|\psi\>$ to be prepared has
$\| \rho_B \|_\infty \leq \smfrac{1}{d}(1+\e)$ and can be prepared
with probability at least $\smfrac{1}{1+\e}$.

Thus, assuming the setting of Lemma 1, \eq{long} is a statement that
the following protocol will succeed with probability at least $1-\e
\over 1+\e$ for {\em all} states $|\psi\>$:
\\[1.5ex]  To send any state $|\psi\>$, 
\\[1.5ex] 1. Alice and Bob draw a random $k \in \{1,\cdots,n\}$ using
$\log n$ bits of shared randomness.  Using \eq{long}, the probability
that $\| \Tr_A U_k |\psi\>\<\psi|
U_k^\dagger \|_\infty < \smfrac{1+\e}{d}$ is at least $1-\e$.
\\[1.5ex] 2. Alice prepares $U_k |\psi\>$ using \eq{prprotocol} and
Bob applies $U_k^\dagger$ to obtain the correct state $|\psi\>$.  With
probability at least $1-\e$, this procedure succeeds with probability
greater than $\smfrac{1}{1+\e}$.
\\[1.5ex] Let $U_k |\psi\> = (X_k \ot I) |\Phi_d\>$.  Note that $X_k$
is a $d\times n$ matrix where $n=(112 \ln 2)(d \log d)/\e^2$.
The entire protocol can be represented by the circuit
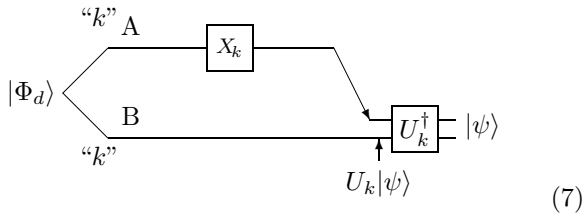
\bea
\setlength{\unitlength}{0.6mm}
\centering
\begin{picture}(90,42)
\put(-7,20){\makebox(10,10){$|\Phi_d\>$}}
\put(5,25){\line(1,1){10}}
\put(5,25){\line(1,-1){10}}
\put(15,35){\makebox(10,10){A}}
\put(15,15){\makebox(10,10){B}}
\put(8,37){\makebox(10,10){``$k$''}}
\put(8,06){\makebox(10,10){``$k$''}}
\put(15,35){\line(1,0){22}}
\put(15,15){\line(1,0){63}}
\put(37,30){\framebox(10,10){\footnotesize{$X_{\!k}$}}}
\put(47,35){\line(1,0){15}}
\put(62,35){\line(1,0){3}}
\put(65,35){\vector(1,-2){8}}
\put(73,19){\line(1,0){5}}
\put(78,12){\framebox(10,10){$U_k^\dagger$}}
\put(88,19){\line(1,0){4}}
\put(88,15){\line(1,0){4}}
\put(93,12.5){\makebox(10,10){$|\psi\>$}}
\put(75,10){\vector(0,1){5}}
\put(70,0){\makebox(10,10){$U_k|\psi\>$}}
\end{picture}
\label{eq:protocol}
\eea
This protocol requires sending system $A$ of dimension 
$\smfrac{112 \ln 2}{\e^2}\, d \log d$  and consuming $log d$ ebits.    
If the goal is to prepare an arbitrary $2l$-qubit state,
we must take $d = 2^l$ and, upon taking logs, we find that the
protocol requires $l + \log l + 2 \log \smfrac{1}{\e} + 7
= l + o(l)$ qubits of communication, $l$ ebits and
$\log n \approx l + \log l + 3 \log \smfrac{1}{\e} + 7 = l + o(l)$
shared random bits. Asymptotically,
this gives a rate of $2$ remote qubits prepared for every qubit
communicated, ebit consumed and random bit shared.
%
%

{\bf Superdense coding of entangled states.~~} Our protocol for
preparing pure states can be easily adapted to enable Alice and Bob
to share a state $|\psi\>$ starting from $|\Phi_d\>$.  We have 
an analog of the probabilistic protocol in \eq{prprotocol}:
\bea
\setlength{\unitlength}{0.6mm}
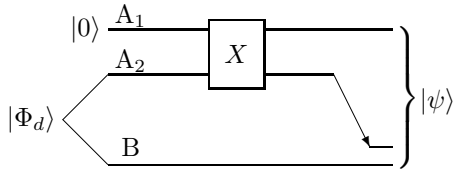
\centering
\begin{picture}(90,42)
\put(-7,10){\makebox(10,10){$|\Phi_d\>$}}
\put(5,15){\line(1,1){10}}
\put(5,15){\line(1,-1){10}}
\put(15,33){\makebox(10,10){A$_1$}}
\put(15,23){\makebox(10,10){A$_2$}}
\put(15,04){\makebox(10,10){B}}
\put(15,25){\line(1,0){20}}
\put(15,05){\line(1,0){63}}
\put(35,25){\line(1,0){2.5}}
\put(37.5,22){\framebox(12,15){$X$}}
\put(49.5,35){\line(1,0){12.5}}
\put(49.5,25){\line(1,0){12.5}}
\put(62,25){\line(1,0){3}}
\put(62,35){\line(1,0){16}}
\put(65,25){\vector(1,-2){8}}
\put(73,09){\line(1,0){5}}
\put(80,02.7){\makebox(3,35){$\!\left.\rule{0pt}{7ex}\right\}$}}
\put(83,14){\makebox(10,10){$|\psi\>$}}
\put(5,30){\makebox(10,10){$|0\>$}}
\put(15,35){\line(1,0){20}}
\put(35,35){\line(1,0){2.5}}
\end{picture}
\non
\eea
where again, $\ket{\psi} = (X_{A_1 A_2} \ot I_B) \, \ket{\Ph_d}$.  The
nonunitary operation $X$ can be performed with probability
$\smfrac{1}{d \, \| \Tr_B \!  |\psi\>\<\psi| \, \|_\infty}$.

Again, the idea is to randomize $|\psi\>$ to make it highly
entangled across the $A_1 A_2$ vs $B$ partition.  However, Bob has no
access to $A_1$ so the randomization operation should only act on
$A_2$ and $B$.  The resulting protocol can be represented by the
circuit
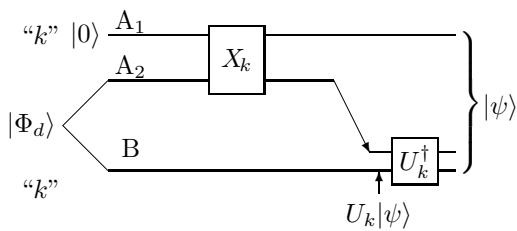
\bea
\setlength{\unitlength}{0.6mm}
\centering
\begin{picture}(90,53)
\put(-7,20){\makebox(10,10){$|\Phi_d\>$}}
\put(5,25){\line(1,1){10}}
\put(5,25){\line(1,-1){10}}
\put(15,43){\makebox(10,10){A$_1$}}
\put(15,33){\makebox(10,10){A$_2$}}
\put(15,15){\makebox(10,10){B}}
\put(-5,40){\makebox(10,10){``$k$''}}
\put(-5,6){\makebox(10,10){``$k$''}}
\put(15,35){\line(1,0){20}}
\put(15,15){\line(1,0){63}}
\put(35,35){\line(1,0){2.5}}
\put(37.5,32){\framebox(12,15){$X_{\!k}$}}
\put(49.5,45){\line(1,0){12.5}}
\put(49.5,35){\line(1,0){12.5}}
\put(62,35){\line(1,0){3}}
\put(62,45){\line(1,0){30}}
\put(65,35){\vector(1,-2){8}}
\put(73,19){\line(1,0){5}}
\put(78,12){\framebox(10,10){$U_k^\dagger$}}
\put(88,19){\line(1,0){4}}
\put(88,15){\line(1,0){4}}
\put(94,12.7){\makebox(3,35){$\!\left.\rule{0pt}{7ex}\right\}$}}
\put(97,24){\makebox(10,10){$|\psi\>$}}
\put(75,10){\vector(0,1){5}}
\put(70,0){\makebox(10,10){$U_k|\psi\>$}}
\put(5,40){\makebox(10,10){$|0\>$}}
\put(15,45){\line(1,0){20}}
\put(35,45){\line(1,0){2.5}}
\end{picture}
\non
\eea
which is the analog of \eq{protocol}.  

The above protocol shares a state with $d^2 = 2^{2l}$ dimensions on
Bob's side.  The quantum communication and entanglement resources
required are the same as in the unentangled case, but more shared
randomness is needed, since the randomization is less effective when
restricted to part of the system.  We prove in
Appendix~\ref{sec:lement} that $\log n \approx 3 l + 2 \log l + 5 \log
\smfrac{1}{\e} + 13$ shared random bits are sufficient.

{\bf Discussion.~~}  
The most intriguing open question is whether shared randomness is
required to perform superdense coding of quantum states.  
Our protocol is universal -- the pure state to be prepared or shared
is completely arbitrary, and no restriction is imposed on its
distribution.
Shared randomness is not needed when the state to be communicated is
drawn from an ensemble, and is negligible when the state is a tensor
product of arbitrary states in blocks of $o(l)$ qubits.
We note that there is a completely different method to achieve
superdense coding of tensor product quantum states (also without shared
randomness) using remote state preparation \cite{Bennett03} and
superdense coding \cite{Bennett92} in the framework of coherent
classical communication \cite{Harrow03a}.  Conversely, teleportation
\cite{Bennett93} and superdense coding of quantum states can be
combined to rederive remote state preparation \cite{Bennett03}, albeit
with a new shared randomness cost.

{\bf Acknowledgements.~~} AH acknowledges funding from the NSA and
ARDA under ARO contract DAAD19-01-1-06. PH and DL acknowledge the
support of the Sherman Fairchild, the Richard Tolman, and the Croucher 
foundations, as well as the US NSF under grant no. EIA-0086038.

\appendix 

\raggedbottom

\section{Proof of Lemma \ref{lem:pure}}
\label{sec:lempure}
Our methods are similar to those detailed in \cite{Hayden03}.  Suppose
each $U_k$ is drawn i.i.d.~according to the Haar (i.e. left and right
unitarily invariant) measure~\cite{Duistermaat99}.
First, fix two arbitrary pure states in ${\cal H}_B$ and ${\cal
H}_{AB}:=\cH_A \ot \cH_B$ and denote their density matrices by $\phi$
and $\psi$.  Also, let $d_A = \dim(\cH_A)$ and $d_B = \dim(\cH_B)$.
Then, for $0 < \e \leq 1$,
\bea
	& & \mPr{U} \lpL \Tr ( \phi \, \Tr_A \! U \psi U^\dag ) \geq 
			\smfrac{1}{d_{\!B}}+\smfrac{\e}{2 \ss d_{\!B}} \rpL
\non
\\
	& = &  \mPr{U} \lpL \Tr ( I \ot \phi) (U \psi U^\dag ) \geq 
			\smfrac{1}{d_{\!B}}+\smfrac{\e}{2 \ss d_{\!B}} \rpL
\non
\\
	& \leq & \exp \left (-\frac{d_{\!A} \e^2}{14 \ln 2} \right) \,, 
\non
\eea
where the last line follows from an argument almost identical to
the proof of Lemma II.3 in \cite{Hayden03}. 
%

Our second step is to prove that
\bea
	& & 
	\mPr{U} \lpL \| \Tr_A \! U \psi U^\dag \|_\infty \geq 
			\smfrac{1}{d_{\!B}} + \smfrac{3\e}{4d_{\!B}} \rpL
\non
\\
	& \leq & \left( {10 \, d_{\!B} \over \e} \right)^{2 \ss d_{\!B}} 
	\exp \! \left( \! -\frac{d_{\!A} \, \e^2}{14 \ln 2} \right) \!.  
\label{eq:second}
\eea

$\cM$ is called a $\delta$-net for pure states in an $m$-dimensional
Hilbert space $\cH$ if, for any pure state $\eta \in \cH$, $\exists
\tilde{\eta} \in \cM$ such that $\|\eta - \tilde{\eta} \|_1 \leq
\delta$.  Here, $\|\cdot\|_1$ denotes the trace norm, which is the sum
of the absolute values of the eigenvalues.  Lemma II.4 in
\cite{Hayden03} states that $\exists \cM$ such that $|\cM| \leq
(\smfrac{5}{\delta})^{2m}$.  We always refer to this type of net in
our discussion.  Note that for any operator $O \in [0,\id]$, $\Tr \lpm
(\eta - \tilde{\eta}) \, O \rpm \leq {\delta \over 2}$.  We will use
this fact often, and we call it ``Fact 1.''

Let $\cM_B$ be an ${\e \over 2d_{\!B}}$-net for pure states in
$\cH_B$.  Then, using Fact 1, 
\bea
	\| \Tr_A U \psi U^\dag \|_\infty 
	& = & \sup_{\phi \in \cH_B} \Tr \lpm \phi \Tr_A U \psi U^\dag \rpm 
\non
\\
	& \leq  & \sup_{\tilde{\phi} \in \cM_B} 
	\Tr \lpm \phi \Tr_A U \tilde{\psi} U^\dag \rpm 
	+ {\e \over 4 d_{\!B}} \,. 
\non
\eea
Thus,
\bea
	& & 
	\mPr{U} \lpL \| \Tr_A \! U \psi U^\dag \|_\infty \geq 
			\smfrac{1}{d_{\!B}} + \smfrac{3\e}{4d_{\!B}} \rpL
\non
\\
	& \leq & 
	\mPr{U} \lpH \; \sup_{\tilde{\phi} \in \cM_{\!B}} \!\!
	\Tr \lpm \tilde{\phi} \Tr_A U \psi U^\dag \rpm \geq 
			\smfrac{1}{d_{\!B}} + \smfrac{\e}{2d_{\!B}} \rpH
\non
\\
	& \leq & \left( {10 \, d_{\!B} \over \e} \! \right)^{\!2 \ss d_{\!B}} 
	\exp \! \left( \! -\frac{d_A \e^2}{14 \ln 2} \right) 
	=: \mu \,,
\label{eq:secondproved}
\eea
where the last line is obtained using the union bound.

In our third step, we introduce the binary random variables  
\bea
	X_k = \left\{ 
	\begin{array}{l} 
	1 ~~{\rm if} ~~ \| \Tr_A U_k \psi U_k^\dagger \|_\infty 
		\geq \smfrac{1}{d_{\!B}} ( 1+\smfrac{3\e}{4} ) \\[1.2ex]
	0 ~~{\rm otherwise.}
	\end{array} \right.
\label{eq:xkdef}
\eea
Then the $X_k$ are i.i.d.~with expectation (over $U_k$) $\EE X_k \leq \mu$
because of \eq{secondproved}.  For a fixed $\psi$, 
\bea
	& \mPr{\{U_k\}} & \!\! \! 
	\lbL {1 \over n} \sum_{k=1}^{n} X_k > \e \rbL 
	\leq \exp \lpm -n D(\e \| \mu) \rpm   
\label{eq:fixpsi}
\\ 
	& & \leq \exp \lpL -n \lpm \e \lpm 
	-2 d_{\!B} \log \smfrac{10 d_{\!B}}{\e} + 
			\smfrac{d_{\!A} \e^2}{14 \ln 2} \rpm - 1 \rpm   \rpL
\non 
\eea
where $D$ is the divergence, with 
\bea
	D(\e \| \mu) & : = & ~\e \log \e + (1-\e) \log (1-\e) 
\non
\\	& &  - \e \log \mu - (1-\e) \log (1-\mu)
\non 
\\	&\geq & -1 - \e \log \mu  \,.
\non
\eea

Consider an $\smfrac{\e}{2d_{\!B}}$-net $\cM_{AB}$ for $|\psi\> \in
\cH_A \ot \cH_B$.  Then,
\bea 
	& \mPr{\{U_k\}} & \! \! \left( \sup_{\tilde{\psi} \in \cM_{\!A\!B}}
	{1 \over n} \sum_{k=1}^n X_K
	> \e \right) 
\label{eq:badeventpure}
\\[1.2ex]
	& \leq & \hspace*{-2.5ex} 
	\left(\! \frac{10 d_{\!B}}{\e} \! \right)^{\!\!2 d_{\!A} d_{\!B}}
	\hspace*{-3ex} 
	\exp \lpL\! -\! n \lpm \e \lpm\! 
	- \!2 d_{\!B} \log \smfrac{10 d_{\!B}}{\e} 
	\!+ \!\smfrac{d_{\!A} \e^2}{14 \ln 2} \rpm - 1 \rpm   \rpL ,
\non
\eea
by the union bound.  If 
\bea
	n > \frac{2 d_{\!A} d_{\!B} \log \smfrac{10 d_{\!B}}{\e}}
	{\smfrac{\e^3 d_{\!A}}{14 \ln 2} 
	- 2 \e \, d_{\!B} \log \smfrac{10 \, d_{\!B}}{\e} -1}
\label{eq:npure}
\eea
the probability in \eq{badeventpure} is strictly less than $1$ and
there exists a choice of $\{U_k\}$ such that the corresponding event
does {\em not} happen.  That is, 
\bea
	\sup_{\tilde{\psi} \in \cM_{\!A\!B}}
	{1 \over n} \sum_{k=1}^n X_K 
	\leq \e  \,.
\eea
Rephrasing the above using \eq{xkdef}, 
\bea
	\forall \tilde{\psi} \in \cM_{\!A\!B} \,, 
	\mPr{k} \lbL \| \Tr_A U_k \tilde{\psi} U_k^\dag \|_\infty \geq 
	\smfrac{1}{d_{\!B}} (1+ \smfrac{3 \e}{4}) \rbL \leq \e \,.
\non
\eea
Finally, applying Fact 1 to the net $\cM_{\!A\!B}$, 
\bea
	\forall \psi \in \cH_{\!A\!B} \,, \exists \tilde{\psi} ~{\rm such~that}~ 
	\hspace*{28ex}
\non
\\
	\| \Tr_A U_k \psi U_k^\dag \|_\infty \leq 
 	\| \Tr_A U_k \tilde{\psi} U_k^\dag \|_\infty 
  	+ \smfrac{\e}{4 d_{\!B}}
\non
\eea
Putting the last two equations together, 
\bea
	\forall \psi  \in \cH_{\!A\!B} \,, 
	\mPr{k} \lbL \| \Tr_A U_k \, \psi \, U_k^\dag \|_\infty \geq 
	\smfrac{1}{d_{\!B}} (1+ \e) \rbL \leq \e \,.
\non
\eea
This completes the proof of Lemma \ref{lem:pure} when we choose 
$d_{\!B} = d$ and $d_{\!A} = \smfrac{112 \ln 2}{\e^2} \, d \log d$. 
With these parameters, and with the hypothesis $d_{\!B} \geq
\smfrac{10}{\e}$, it is straightforward to verify that $\smfrac{120 \ln
2}{\e^3} \, d \log d$ is an upper bound to the required number of
unitaries in \eq{npure}.
%

\section{Extension of lemma \ref{lem:pure}   for sharing entangled states}
\label{sec:lement}

We want to randomize $|\psi\>$ to make it highly entangled across the
$A_1 A_2$ vs $B$ partition.  However, Bob has no access to $A_1$ so
the randomization operation should only act on $A_2$ and $B$.
How fast the probability of success approaches $1$, as a function of
$\dim(A_2)$ and the number of randomizing operations is given by the
following analog of Lemma \ref{lem:pure}:
\begin{lemma} \label{lem:ent}
Let $0 < \e \leq 1$.  If $d \geq \smfrac{10}{\e}$, 
$\exists \{U_k\}_{k=1}^{n}$, where $n = \smfrac{13440 (\ln
2)^{2\!\!}}{\e^5} \; d^3 (\log d)^2$, such that
\bea
	\forall |\psi\>,  
	& \mPr{k} & \!\!\! \left( 
 	\|\! \Tr_{\!A_1 \!A_2} I_{\! A_1 \!} \! \ot \! U_k \, |\psi\>\<\psi| 
	\, I_{\! A_1 \!} \! \ot \! U_k^\dagger \, \|_\infty \! < \!
        \smfrac{1+\epsilon}{d} \right)  \!\!
\non
\\
	& \geq & \!\! 1\!-\!\e \,. 
\label{eq:longent}
\eea
Here, each $U_k$ takes $d^2$-dimensional states into a Hilbert space
${\cal H}_{A_2} \ot {\cal H}_B$ where $\dim(\cH_{A_2}) = \smfrac{112 \ln
2}{\e^2} \, d \log d$ and $\dim(\cH_B) = d$.
%
%
\end{lemma}
\eq{longent} ensures that with probability at least $1-\e$ (over $k$),
the state is shared with probability greater than $\smfrac{1}{1+\e}$,
so the overall probability of success is at least
$\smfrac{1-\e}{1+\e}$.

The proof is almost the same as that of Lemma \ref{lem:pure} and we
only outline the differences, leaving out details that are already
discussed.  We select the $U_k$ independently and identically 
distributed (i.i.d.)~according to the Haar measure as before, and
choose $d_{A_1 \!}  := \dim(\cH_{A_1})$, $d_{A_2 \!} :=
\dim(\cH_{A_2})$, and $d_{B \!} := \dim(\cH_B)$ later.  Let $\psi$ be
a fixed pure state in $\cH_{A_1 A_2 B} := \cH_{A_1} \ot \cH_{A_2} \ot
\cH_B$.  By the convexity of $\| \cdot \|_\infty $,
\bea
	& & \| \Tr_{A_1 A_2} 
	I_{A_1}\!\! \ot\! U_k \, \psi \, I_{A_1} \!\!\ot\! U_k^\dag \|_\infty
\non
\\
	& = & \| \Tr_{A_2} 
	U_k \,  (\Tr_{A_1} \psi) \, U_k^\dag \|_\infty
\non
\\
	& \leq & \max_j \| \Tr_{A_2} 
	U_k \,  \eta_j \, U_k^\dag \|_\infty
\eea
where $\{\eta_j\}$ are the eigenvectors of $\Tr_{A_1} \psi$.   
Thus, 
\bea
	& & \mPr{\{U_k\}} \left( \| \Tr_{A_1 A_2} 
	I_{A_1}\!\! \ot\! U_k \, \psi \, I_{A_1} \!\!\ot\! U_k^\dag \|_\infty
	\geq \smfrac{1}{d_B} + \smfrac{\e}{2d_B} \right)
\non
\\
	& \leq & \mPr{\{U_k\}} \left( 
	\max_j \| \Tr_{A_2} U_k \,  \eta_j \, U_k^\dag \|_\infty
	\geq \smfrac{1}{d_B} + \smfrac{\e}{2d_B} \right)
\non
\\
	& \leq & \left( {10 \, d_{\!B} \over \e} \right)^{2 \ss d_{\!B}} 
	\exp \! \left( \! -\frac{d_{\!{A_2}} \, \e^2}{14 \ln 2} \right)   
	:= \mu \,, 
\non
\eea
by applying \eq{second} with $d_{A}$ replaced by $d_{A_2}$.  
We similarly define the binary random variables 
\bea
	X_k = \left\{ 
	\begin{array}{l} 
	1 ~~{\rm if} ~~ \| \Tr_{A_1  \! A_2} I_{A_1}  \!  \! \ot  \! U_k 
			\, \psi \, I_{A_1} \!\! \ot \! U_k^\dagger \|_\infty 
		\geq \smfrac{1}{d_{\!B}} ( 1\!+\!\smfrac{3\e}{4} ) \\[1.2ex]
	0 ~~{\rm otherwise.}
	\end{array} \right.
\non
\eea
\eq{fixpsi} still holds here: 
\bea
	& \mPr{\{U_k\}} & \!\! \! 
	\lbL {1 \over n} \sum_{k=1}^{n} X_k > \e \rbL 
\non
\\ 
	& & \leq \exp \lpL -n \lpm \e \lpm 
	-2 d_{\!B} \log \smfrac{10 d_{\!B}}{\e} + 
			\smfrac{d_{\!A_2} \e^2}{14 \ln 2} \rpm - 1 \rpm \rpL
\non 
\eea
Again, we choose an $\smfrac{\e}{2d_{\!B}}$-net $\cM_{A_1 A_2 B}$ 
for $\psi \in \cH_{A_1 A_2 B}$, so that 
\bea 
	\!\!\!\!\!
	& \mPr{\{U_k\}} & \! \! \left( \sup_{\tilde{\psi} \in \cM_{\!A\!B}}
	{1 \over n} \sum_{k=1}^n X_K
	> \e \right) 
\non
\\[1.2ex]
	\!\!\!\!\!
	& \leq & \hspace*{-2.5ex} 
	\left(\! \frac{10 d_{\!B}}{\e} \! \right)^{\!\!2 d_{\!A_{\!1}}
	 \ms d_{\!A_{\!2}} 
	 \ms d_{\!B}}
	\hspace*{-5ex} 
	\exp \lpL\! -\! n \lpm \e \lpm\! 
	- \!2 d_{\!B} \log \smfrac{10 d_{\!B}}{\e} 
	\!+ \!\smfrac{d_{\!A_2} \e^2}{14 \ln 2} \rpm - 1 \rpm   \rpL \! .
\non
\eea
{From} previous analysis, the choice $d_{A_2} = \smfrac{112 \ln
2}{\e^2} d \log d$, $d_{B} = d$ and $n \geq d_{A_1} \smfrac{120 \ln
2}{\e^3} \, d \log d$ is sufficient to bound the above probability 
away from $1$, in which case there exists $\{U_k\}$ for which
\bea
	\sup_{\tilde{\psi} \in \cM_{\!A\!B}}
	{1 \over n} \sum_{k=1}^n X_K
	\leq \e \,, 
\eea
and 
\bea
	& \forall \psi &  \in \cH_{\!A_1\!A_2 \!B} \,, 
\non
\\
	& \mPr{k} & \lbL \| \Tr_{A_1 A_2} I_{A_1} \!\!\ot \!U_k \, \psi \, 
	I_{A_1} \!\! \ot \! U_k^\dag \|_\infty \geq 
	\smfrac{1}{d_{\!B}} (1+ \e) \rbL \leq \e \,.
\non
\eea
Even though $d_{A_1}$ is arbitrary, by a local change of basis on
$A_1$ alone (which can always be done without any resource after the
sharing), one can assume $\dim(A_1) \leq \dim(A_2) \dim(B)$.  This
means $n = \smfrac{13440 (\ln 2)^{2\!}}{\e^5} \; d^3 (\log d)^2$ is
sufficient.

\clearpage 


\end{document}